\documentclass[twoside,twocolumn,english,pra,twocolumn,aps,superscriptaddress]{revtex4}
\usepackage{amsmath}
\usepackage{amsthm}
\usepackage{graphicx}
\usepackage[unicode=true,pdfusetitle,
 bookmarks=true,bookmarksnumbered=false,bookmarksopen=false,
 breaklinks=false,pdfborder={0 0 0},backref=false,colorlinks=true]
 {hyperref}
\usepackage{breakurl}
\usepackage{epstopdf}
\usepackage{amsmath,bm}

\newcommand{\ehbar}{\hbar_{\mathrm{eff}}}

\begin{document}

\title{Ballistic diffusion vs. damped oscillation of energy in a $\mathcal{PT}$-symmetric quantum kicked harmonic oscillator}

\author{Jian-Zheng Li}
\email{lijianzheng@jxust.edu.cn}
\affiliation{School of Science, Jiangxi University of Science and Technology, Ganzhou 341000, China}

\begin{abstract}
We numerically study the quantum dynamics of a $\mathcal{PT}$-symmetric kicked harmonic oscillator. We observe that directed current of momentum and ballistic diffusion of energy coexist under the non-resonant conditions, whereas both the momentum and energy oscillate as damped cosine functions with identical frequencies under the resonant conditions. The research shows that the directed current of momentum and ballistic diffusion of energy arise from nearest-neighbor hopping between momentum eigenstates with the non-Hermitian driving, while the damped oscillations of momentum and energy originate from resonant coupling between the non-Hermitian driving and the harmonic oscillator. Our findings indicate that the non-Hermiticity and the frequency characteristic of this system collectively result in these distinctive dynamical behaviors.
\end{abstract}
\date{\today}
\maketitle

\section{Introduction}\label{intro}
Recently, quantum dynamics of Floquet-driven systems~\cite{ZHuang21,CFleckenstein21} has been one of hot research topics in many areas, such as quantum optics~\cite{CMa18,YZhang23}, solid state electronics~\cite{ADiazFernandez19,AFAdiyatullin23}, cold atom physics~\cite{MReitter17,JHSToh22,SSMaurya22}, quantum circuits~\cite{FPetiziol22,KSu22} and so on. The kicked harmonic oscillator (KHO) is one of the most important Floquet-driven models~\cite{GAKells04,EGArrais16,SWimberger05,JHSToh22,MSCustodio17}. This system can be regarded as not only a model of charged particles kicked in a uniform magnetic field but also a model of electronic transport in semiconductor supper-lattices~\cite{IDana95,SGupta11,TMFromhold01}. The quantum dynamics of this system depends dramatically on system parameters, such as the ratio of the oscillation frequency ($\omega$) to the kicking frequency ($\Omega$), i.e., $\omega/\Omega=R$, and the kicking strength ~\cite{GAKells04,JHSToh22,NDVarikuti24}. The ratio $R$ determines whether the system is in the resonance state~\cite{AAChernikov87}. Specifically, the resonance occurs when $R$ is rational, whereas the non-resonance is characterized by an irrational number $R$. Furthermore, the enhancement of kicking strength facilitates the transition from dynamical localization to delocalization.~\cite{GAKells04}. So the introduction of one new system parameter will has a significant influence on the quantum dynamics of KHO.

It is well established that a significant modification to conventional quantum mechanics involves the introduction of non-Hermitian Hamiltonian~\cite{REGanainy18,Ashida20,SFranca22}. It can describe an open system that exchanges energy or particles with its surrounding environment, such as photonic crystals with complex dielectric constants~\cite{KDing15,MLNChen21,WZhu23}, ultracold atoms in complex potential fields~\cite{MKreibich14,PHe21}, and ferromagnetics with complex spin coupling~\cite{XZhang24}. In recent years, the study of quantum dynamics in non-Hermitian systems has garnered significant attention. The skin effect and bulk-boundary correspondence are found in the non-Hermitian Su-Schrieffer-Heeger (SSH) system with the Floquet driving~\cite{KShi24}. In the non-Hermitian quantum kicked rotor (QKR) system which is also a paradigmatic Floquet model, the non-Hermitian kicking potential can enhance the effect of dynamical localization~\cite{KQHuang21}. The $\cal{PT}$-symmetric~\cite{KGMakris08,CERuter10} non-Hermitian kicking potential in QKR can result in directed momentum current and ballistic energy diffusion~\cite{WLZhao19,JZLi23}. However, the dynamical behaviors of momentum and energy in the non-Hermitian systems remains many unresolved questions. In this context, the quantum dynamical behaviors originating from the non-Hermiticity in the KHO model require immediate investigation.

In this paper, we study the quantum dynamical evolutions of momentum and energy in a $\cal{PT}$-symmetric non-Hermitian KHO system under the non-resonant and resonant conditions. Under the non-resonant conditions, the directed current of momentum and ballistic diffusion of energy coexist in this system. The harmonic oscillator system can be seen as a quasi-classical particle according to its probability density distributions in both momentum and coordinate spaces. Under the resonant conditions, we observe damped oscillations of momentum and energy in this system. The dynamical behaviors of momentum and energy are described by damped cosine functions with identical frequencies, so the system is regarded as a damped harmonic oscillator. Our investigation suggests the interactions of the non-Hermitian driving and the harmonic oscillator are responsible for these distinctive dynamical behaviors under the non-resonant and resonant conditions.

This paper is organized as follows. In Sec.~\ref{SEC-MD}, we describe the system and carry out the non-dimensionalization. In Sec.~\ref{Dynbehavior}, we show the dynamical behaviors of momentum and energy in our system under both the non-resonant and resonant conditions. A summary is presented in Sec.~\ref{SEC-SUM}.
%%%%
\section{Physical Model}\label{SEC-MD}
We consider the $\cal{PT}$-symmetric quantum kicked harmonic oscillator (PTQKHO) system, which is described by the Hamiltonian
\begin{equation}\label{Hamil}
  {\cal{H}} =\frac{P^2}{2M}+\frac{M\omega^2 X^2}{2}+V(X) \sum^\infty_{n=0}\delta(t^{'}-nT),
\end{equation}
where $P$ is the momentum operator, $X$ is the coordinate, $M$ is the mass, $\omega$ is the oscillation frequency of the harmonic oscillator, and $T$ is the time interval between two kicks. In our system, specially, the harmonic potential ${\cal{H}}_{p}=M \omega^2 X^2/2$ is a periodic function, owning the same period with the kicking potential $V(X)=V_{0}[\cos (\frac{2\pi}{b}X)+i \lambda \sin (\frac{2\pi}{b}X)]$, i.e., ${\cal{H}}_{p}(X+b)={\cal{H}}_{p}(X)$. According to the non-dimensionalization method~\cite{GLemarie09}, we define dimensionless variables as $\theta=2\pi X/b$, $\hbar_{\rm{eff}}=4 \pi^{2} T \hbar/Mb^{2}$, $p=2 \pi T P /Mb$, $\eta=T \omega$, $t=t^{'}/T$, $K=4 \pi^{2} T V_{0} /Mb^{2}$, and $\text{H}=4 \pi^{2} T^{2} {\cal{H}}/Mb^{2}$, so the effective Hamiltonian of this system can be expressed as

\begin{equation}\label{Hamil}
  \text{H}=\frac{p^2}{2}+\frac{\eta^2\theta^2}{2}+K(\cos \theta+i \lambda \sin \theta) \sum^\infty_{n=0}\delta(t-n).
\end{equation}
Here, the coordinate $\theta$ and the momentum operator $p$ obey the commutation relation $[\theta,p]=i\hbar_{\rm{eff}}$. The parameter $\eta$ equals $2\pi \cdot (\omega/\Omega)$ with $\Omega$ being the kicking frequency. The parameter $K$ and $\lambda$ are real part and imaginary part of the kicking strength, respectively. The time $t=n(=0,1,2,\ldots)$ indicates the number of kicks.

The eigenequation of $p$ has the expression $p|\varphi_{m}\rangle=p_{m}|\varphi_{m}\rangle$, with the eigenvalue $p_{m}=m\hbar_{\rm{eff}}$, the eigenstate $\langle \theta|\varphi_{m}\rangle=e^{im\theta}/\sqrt{2\pi}$, and $m=\ldots,-1,0,1,\ldots$. An arbitrary quantum state $|\psi\rangle$ can be expanded on the basis of the state $|\varphi_{m}\rangle$, i.e., $|\psi\rangle=\sum_{m}\psi_{m}|\varphi_{m}\rangle$. In one period from $t=n$ to $t=n+1$, the evolution of the quantum state is given by $|\psi(t=n+1)\rangle=U|\psi(t=n)\rangle$, with the Floquet operator
\begin{equation}
  U=U_{\rm_{\omega}}U_{\rm_{K}}\;,
\end{equation}
where the harmonic oscillator evolution operator takes the form
\begin{equation}
  U_{\rm_{\omega}}=\exp\left[-\frac{i}{\hbar_{\rm_{eff}}}(\frac{p^2}{2}+\frac{\eta^{2}\theta^{2}}{2})\right]\;,
\end{equation}
and the kicking evolution operator is written as
\begin{equation}
  U_{\rm_{K}}=\exp \left[-\frac{i}{\hbar_{\rm_{eff}}}K(\cos \theta+i \lambda \sin \theta)\right]\;.
\end{equation}
In our numerical simulation, the initial state is set to be the ground state of the momentum operator, i.e., $|\psi(t=0)\rangle=|\varphi_{0}\rangle$. The numerical simulation in one evolution period splits into two steps, namely, the kicking evolution and the harmonic oscillator evolution. First, the kicking evolution is conducted, i.e., $|\psi'\rangle=U_K|\psi(t=n)\rangle$. Then, the harmonic oscillator evolution is conducted, i.e., $|\psi(t=n+1)\rangle=U_{\omega}|\psi'\rangle$. According to the split-steps method~\cite{GAKells04,WLZhao14}, the harmonic oscillator operator is expressed in the form below
\begin{equation}
  U_{\omega}=\prod_{j=1}^{N}\left[\text{e}^{-\frac{i}{\hbar_{\text{eff}}} \cdot \frac{p^{2}}{4} \cdot \Delta t_{j}} \cdot \text{e}^{-\frac{i}{\hbar_{\text{eff}}} \cdot \frac{\eta^{2} \theta^{2}}{2} \cdot \Delta t_{j}} \cdot \text{e}^{-\frac{i}{\hbar_{\text{eff}}} \cdot \frac{p^{2}}{4} \cdot \Delta t_{j}}\right],
\end{equation}
with $\Delta t_{j}=1/N$. With such a method, the quantum state at arbitrary time can be given easily.

We investigate the non-Hermitian effects (i.e., $\lambda$) on the quantum dynamics under the non-resonant and resonant conditions by calculating mean momentum $\langle p\rangle=\rm{Tr}(\rho\emph{p})$, kinetic energy $\langle E_{\text{k}}\rangle=\frac{1}{2}\langle p^{2} \rangle=\frac{1}{2}\rm{Tr}(\rho\emph{p}^2)$, potential energy $\langle E_{\text{p}}\rangle=\frac{1}{2} \eta^2 \cdot \rm{Tr}(\rho \theta^2)$, and total energy $\langle E\rangle=\langle E_{\text{k}}\rangle+\langle E_{\text{p}}\rangle$. The norm $\cal{N}$ of quantum state rises sharply with time in the non-Hermitian phase, so we normalize the quantum state to be unity after each kick in order to reduce the norm's contribution to the observable. Thus we define the density matrix of the system as $\rho=\frac{1}{\mathcal{N}}\rm{Tr}(|\psi\rangle\langle\psi|)$.

\section{Results and disscusion}\label{Dynbehavior}

\subsection{Directed current of momentum and ballistic diffusion of energy under non-resonant conditions}
\begin{figure}[htbp]
\begin{center}
\includegraphics[width=8.5cm]{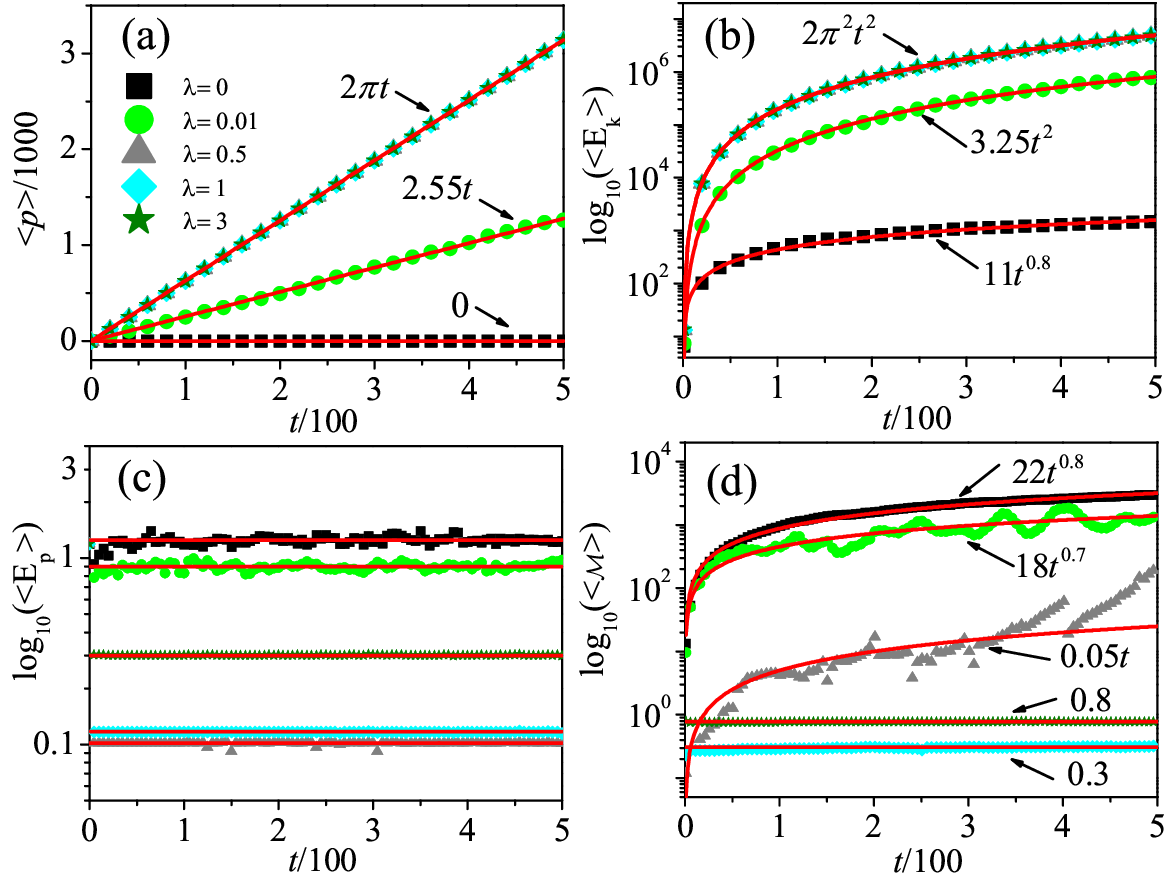}
\caption{Time evolution of $\langle p\rangle$ (a), $\langle E_{k}\rangle$ (b), $\langle E_{p}\rangle$ (c), and $\mathcal{M}$ (d) with $\lambda=0$ (squares), 0.01 (circles), 0.5 (triangles), 1 (diamonds), and 3 (pentagrams). Red lines in (a), (b), (c) and (d) are the fitting functions accordingly. The parameters are $K=5$, $\ehbar=0.1$, and $\eta=2\pi/e^2$.\label{non-resonant Diffusion}}
\end{center}
\end{figure}

Under the non-resonant conditions, e.g., $\eta=2\pi/e^2$, the quantum dynamical behaviors of momentum and energy in the system are discussed in this section. In the Hermitian phase (i.e., $\lambda=0$),  there is no momentum current, i.e., $\langle p\rangle=0$ [see Fig.~\ref{non-resonant Diffusion}(a)], a sub-diffusion behavior of kinetic energy appears, i.e., $\langle E_{\text{k}}\rangle \propto t^{0.8}$ [see Fig.~\ref{non-resonant Diffusion}(b)], and the potential energy doesn't change with time, i.e., $\langle E_{\text{p}}\rangle = C$ [see Fig.~\ref{non-resonant Diffusion}(c)]. Correspondingly, the width of wave package, i.e., $\mathcal{M}= \langle p^2\rangle-\langle p\rangle^{2}$, also increases as a power-exponential function, i.e.,  $\mathcal{M} \propto t^{0.8}$ [see Fig.~\ref{non-resonant Diffusion}(d)], which indicates the appearance of delocalization~\cite{NCherroret14,LErmann14}. In the non-Hermitian phase (i.e., $\lambda \neq 0$), exotic diffusion behaviors are found.
The momentum of the system grows linearly with time,
\begin{equation}\label{LinCurret}
  \langle p \rangle=Gt\;,
\end{equation}
which is the characteristic feature of DC [e.g., $\lambda=1$ in Fig.~\ref{non-resonant Diffusion}(a)]. Moreover,  the growth rate $G$ increases to $2\pi$ with the increase of non-Hermitian driving strength $\lambda$. Meanwhile, the kinetic energy increases with a ballistic function of time [e.g., $\lambda=1$ in Fig.~\ref{non-resonant Diffusion}(b)], i.e., $\langle E_{\text{k}} \rangle\approx \frac{1}{2}G^2t^{2}$.
And the potential energy remains constant throughout the time evolution [e.g., $\lambda=1$ in Fig.~\ref{non-resonant Diffusion}(c)], i.e., $\langle E_{\text{p}}\rangle = C$.
Interestingly, the potential energy decreases firstly and then increases with the increase of $\lambda$. It is evident that the total energy of the system, i.e., $\langle E \rangle = \langle E_{\text{k}}\rangle+\langle E_{\text{p}}\rangle$, can be written as
%%%
\begin{equation}\label{Ballenergy}
  \langle E \rangle \approx \frac{1}{2}G^2t^{2}+C\;,
\end{equation}
which remains a ballistic function of time, with the addition of a constant term.
%%%
Correspondingly, the width of the wave packet increases as a power-exponential function of time, i.e., $\mathcal{M}=\beta t^{\alpha}$ [See Fig.~\ref{non-resonant Diffusion}(d)]. When the driving strength $\lambda$ increases sufficiently (e.g., $\lambda=1,3$), $\mathcal{M}$ approaches a constant, i.e., $\mathcal{M} \propto C$. This indicates that the wave packet exhibits no expansion during its time evolution.
%%%
%%%

We find obviously that, except the constant $\langle E_{p}\rangle$, the evolutions of $\langle p\rangle$, $\langle E_{k}\rangle$, and $\mathcal{M}$ in this system are similar to those of the QKR system~\cite{WLZhao19,JZLi23} respectively in the non-Hermitian phase. According to the mapping method in the references~\cite{ARussomanno21,FMSurace19}, therefore, the effective Hamiltonian of the PTQKHO system under the non-resonant conditions is written as
\begin{equation}\label{Hamil}
\begin{aligned}
  \text{H}_{a}&=\frac{\hbar_{eff}^{2}}{2}\sum^{+\infty}_{m=-\infty} m^2 n_{m}+C \\
  &+\frac{K}{2}[\sum^{+\infty}_{m=-\infty}(a^{\dagger}_{m} a_{m+1}+a^{\dagger}_{m+1} a_{m})\\ &+i \lambda \sum^{+\infty}_{m=-\infty}(a^{\dagger}_{m} a_{m+1}-a^{\dagger}_{m+1} a_{m})] \sum^\infty_{n=0}\delta(t-n).
\end{aligned}
\end{equation}
Here $a_{m}$ and $a_{m}^{\dag}$ are respectively the annihilation and creation operators in the number representation of momentum. The eigenstates $\{| \varphi_{m} \rangle \}$ of the momentum operator $p$ with eigenvalues $m\hbar_{eff}$ is the basis in this representation. $a_{m}$ and $a_{m}^{\dag}$ are bosonic operators which obey the commutation relations $[a_{m}, a_{l}^{\dag}]=\delta_{m,l}$ and  $[a_{m}, a_{l}]=0$. $n_{m}=a_{m}^{\dag} a_{m} $ are the number operators obeying the constraint $\sum_{m} n_{m}P_{m}=1$ where $P_{m}$ is the momentum eigenstate occupancy. $a_{m}^{\dag} a_{m+1}$ is the nearest-neighbor hopping operator from $| \varphi_{m+1} \rangle $ to $| \varphi_{m} \rangle$, and $a_{m+1}^{\dag} a_{m}$ is the opposite operator. It can be concluded that the directed current of momentum and the ballistic diffusion of energy originate from the nearest-neighbor hopping between the momentum eigenstates with the non-Hermitian driving.

%%%%
\begin{figure}[htbp]
\begin{center}
\includegraphics[width=8.5cm]{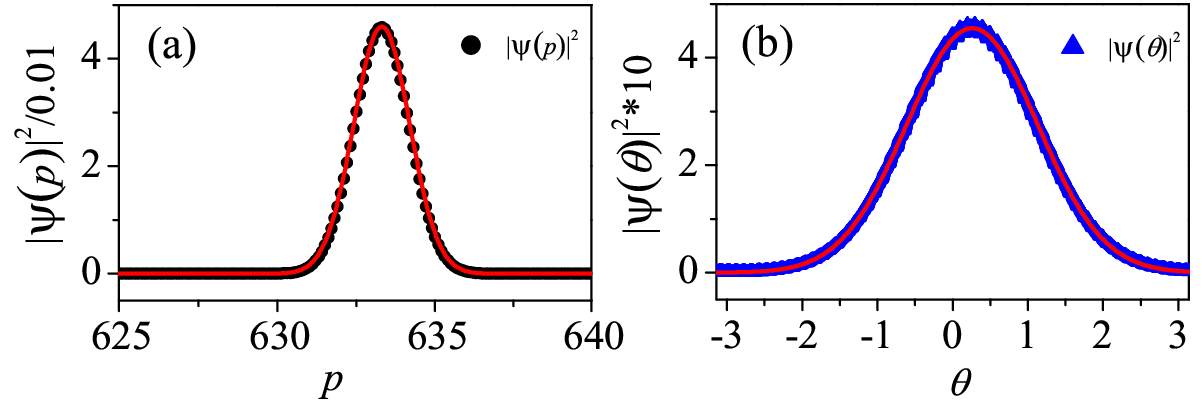}
\caption{(a) Probability density distribution $|\psi(p)|^{2}$ (cicles) in momentum space and (b) probability density distribution $|\psi(\theta)|^{2}$ (triangles) in coordinate space for $t=101$ with $\lambda=3$.  Red lines in (a) and (b) indicate the fitting functions of the Gaussian forms $|\psi(p)|^2\propto e^{-(p-p_c)^2/\sigma_{1}}$ and $|\psi(\theta)|^2\propto e^{-(\theta-\theta_c)^2/\sigma_{2}}$ respectively. Other parameters are consistent with those presented in Fig.~\ref{non-resonant Diffusion}.\label{Wavefunction}}
\end{center}
\end{figure}

The probability density distributions of the system in momentum space and in coordinate space are investigated further. As shown in Fig.~\ref{Wavefunction}(a), the probability density distribution with $\lambda=3$ in momentum space is well described by a Gaussian function, i.e., $|\psi(p,t)|^{2}\propto e^{-[p-p_{c}(t)]^{2}/\sigma_{1}}$, which localizes around $p_{c}(t)(=\langle p\rangle)$. It is noting that the center $p_{c}(t)$ of wave packet moves forward with time, indicating the emergence of the DC in momentum space, too. In addition, its width $\sigma_{1}$ doesn't vary with time, corresponding to the invariance of $\mathcal{M}$. Fig.~\ref{Wavefunction}(b) shows the probability density distribution of the system in coordinate space. We find that it can be well described by another Gaussian function, i.e., $|\psi(\theta,t)|^{2}\propto e^{-[\theta-\theta_{c}(t)]^{2}/\sigma_{2}}$, which spreads throughout the coordinate space from $-\pi$ to $\pi$.
The features of probability density distributions in momentum space and in coordinate space clearly indicate that the system can be seen as a quasi-classical particle under the non-resonant conditions. According to the theory of quasi-classical motion~\cite{FGori00}, the force ($f$) imposed on the particle can be expressed as
\begin{equation}\label{VARiance}
  f= \frac{dp_{c}(t)}{dt}=G\;,
\end{equation}
which is a constant force from the non-Hermitian driving potential field on every kick.
Thus the particle can obtain the directed momentum and ballistic energy successively with the non-Hermitian driving.
\subsection{Oscillations of momentum and energy under resonant conditions}
\begin{figure}[htbp]
\begin{center}
\includegraphics[width=7.5cm]{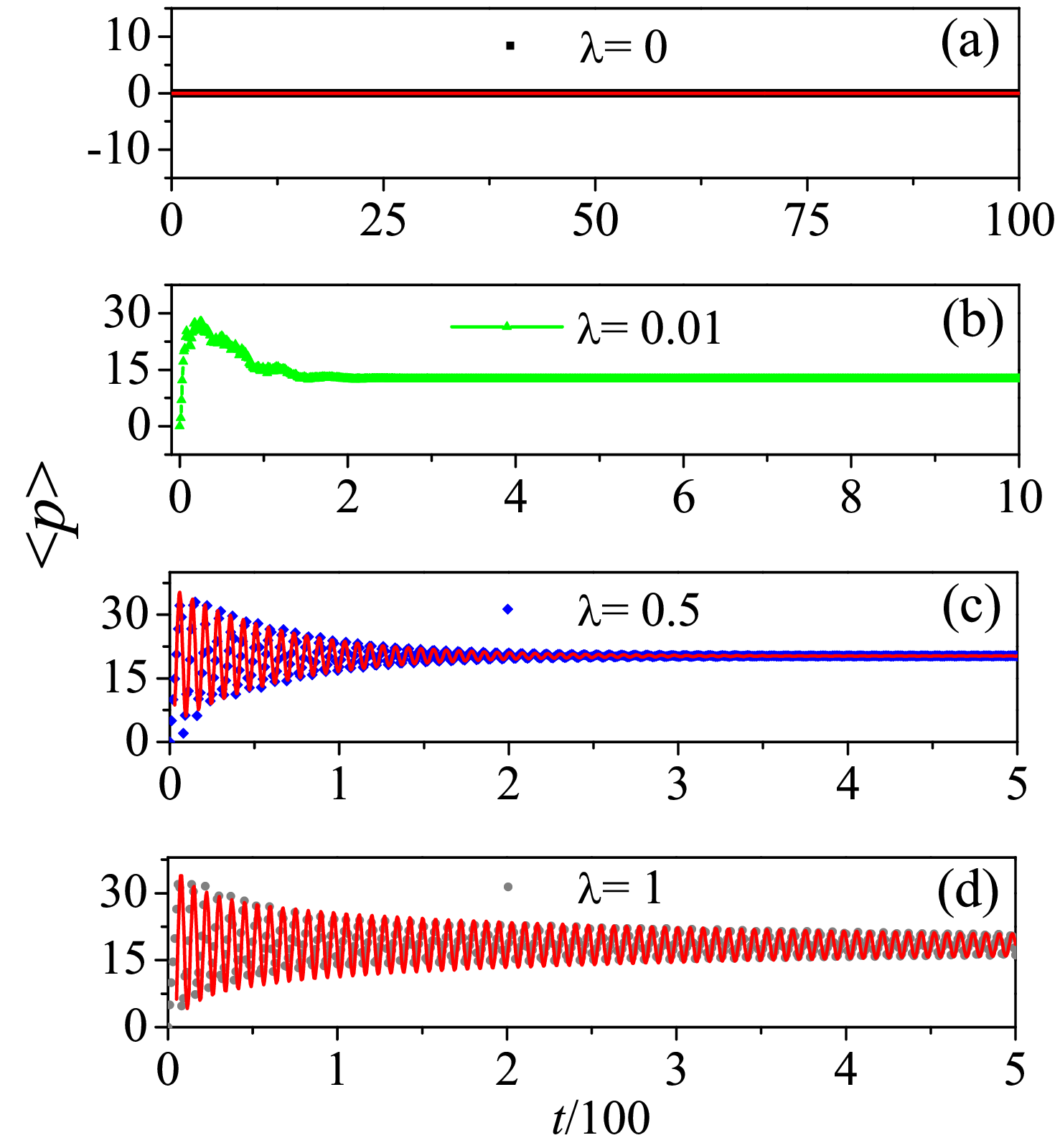}
\caption{Time evolution of $\langle p\rangle$ with $\lambda=0$ (squares) (a), 0.01 (triangles) (b), 0.5 (diamonds) (c), and 1 (circles) (d). Red line in (a) indicates $\langle p\rangle= 0$, and red lines in (c) and (d) indicate $\langle p \rangle=p_{s}-p_{am}(t)\cdot\text{cos}\left[\omega_{c}(t-t_{c})\right]$ in Eq.~\eqref{momentum} with $p_{am}(t)=16.5 \cdot \exp(-\frac{t}{66})$ and $p_{am}(t)=30\cdot \exp\left(-\frac{t}{600} \right)\cdot(\frac{2}{\pi t})^{1/4}$ respectively. $\omega_{c}=\frac{4\pi}{15}$ and $t_{c}=t_{0}+D \cdot \exp(0.01t)$. The parameters are $K=5$, $\ehbar=0.1$, and $\eta=2\pi$.\label{momentum oscillation}}
\end{center}
\end{figure}

In this section, we discuss the dynamical behaviors of momentum and energy in the system under the resonant conditions, e.g., $\eta= 2\pi$. In the Hermitian phase (i.e., $\lambda=0$), there is no oscillation of momentum, i.e., $\langle p\rangle=0$ [see Fig.~\ref{momentum oscillation}(a)]. Interestingly, the oscillation of momentum occurs in the non-Hermitian phase (i.e., $\lambda \neq 0$). When the strength of non-Hermitian driving increases sufficiently, e.g., $\lambda=0.5$ [see Fig.~\ref{momentum oscillation}(c)], the momentum in the system oscillates as a damped cosine function,
\begin{equation}\label{momentum}
  \langle p \rangle=p_{s}-p_{am}(t)\cdot\text{cos}\left[\omega_{c}(t-t_{c})\right]\;,
\end{equation}
with the saturation momentum $p_{s}$, the oscillation frequency $\omega_{c}$, the phase shift $t_{c}=t_{0}+D \cdot \exp(\gamma t)$, and the damped amplitude $p_{am}(t)\propto \exp(-\frac{t}{\tau})$. Continuing to increase the driving strength, e.g., $\lambda=1$ [see Fig.~\ref{momentum oscillation}(d)], the momentum also oscillates as the cosine function, but the amplitude $p_{am}(t)$ decays as another function, i.e., $p_{am}(t)\propto \exp\left(-\frac{t}{\tau}\right)\cdot(\frac{2}{\pi t})^{1/4}$. The research results show that the non-Hermitian driving gives rise to the damped oscillation of momentum and can regulate the amplitude, the phase shift, and even the saturation under the resonant conditions. It provides a pathway to the experimental engineering of quantum dynamic behaviors in momentum space~\cite{BGadway13,DXie20}.

%%%%%%%
\begin{figure}[htbp]
\begin{center}
\includegraphics[width=7.8cm]{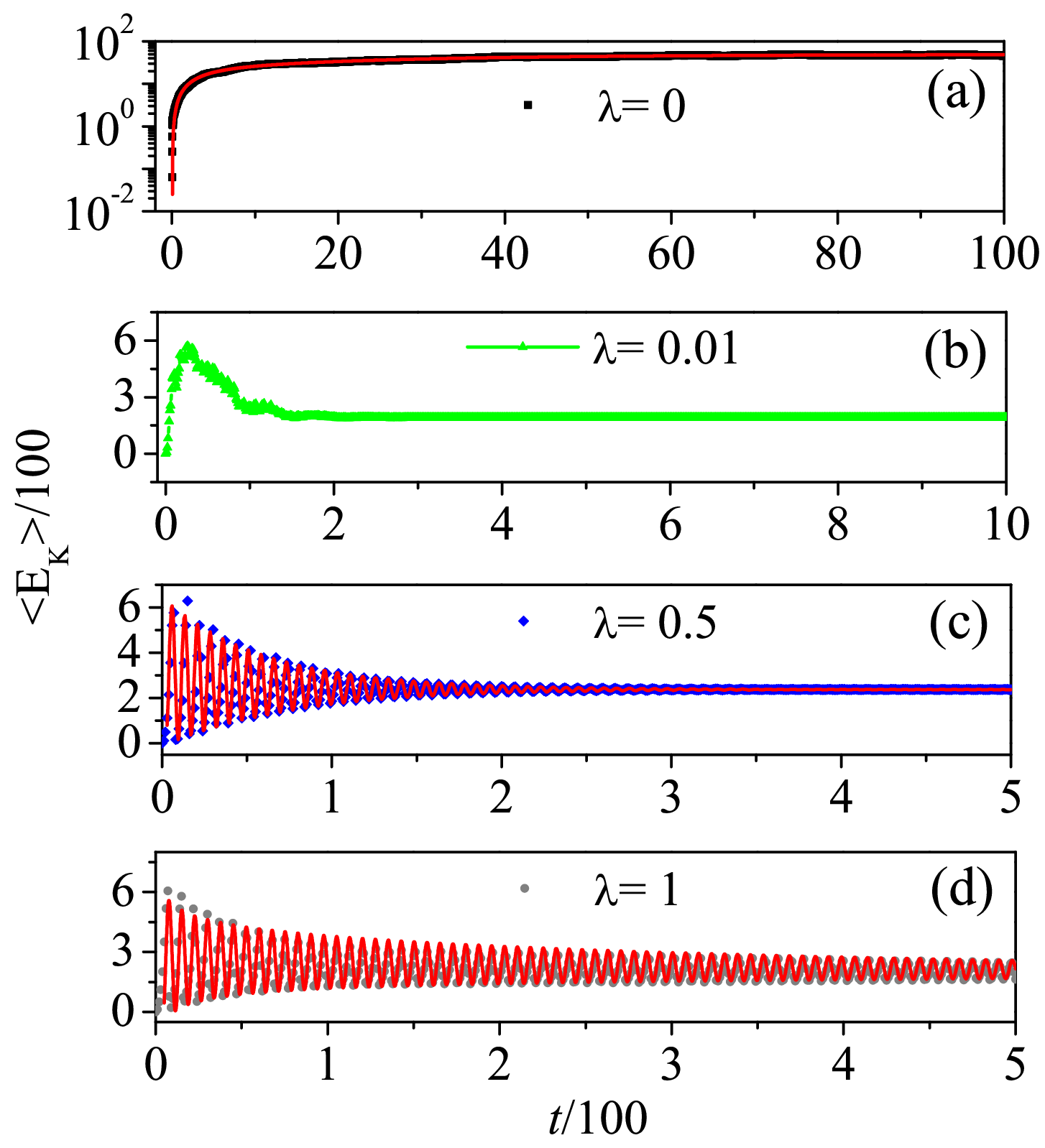}
\caption{Time evolution of $\langle E_{k} \rangle$ with $\lambda=0$ (squares) (a), 0.01 (triangles) (b), 0.5 (diamonds) (c), and 1 (circles) (d). The red line in (a) indicates $\langle E_{k}\rangle=E_{ks}-A_{1}\exp(-\frac{t}{323})-A_{2}\exp(-\frac{t}{2730})$; in (c) $\langle E_{k} \rangle=E_{ka}(t)-E_{km}(t) \cdot \text{cos}\left[\omega_{c}(t-t_{c})\right]$ in Eq.~\eqref{kinetic} with $E_{ka}(t)=E_{ks}+80\cdot\exp(-\frac{t}{40})$ and $E_{km}(t)=330\cdot \exp(-\frac{t}{66})$; in (d) $\langle E_{k} \rangle=E_{ka}(t)-E_{km}(t) \cdot \text{cos}\left[\omega_{c}(t-t_{c})\right]$ in Eq.~\eqref{kinetic} with $E_{ka}(t)=E_{ks}+70\cdot\exp(-\frac{t}{300})$ and $E_{km}(t)=550\cdot \exp\left(-\frac{t}{600}\right)(\frac{2}{\pi t})^{1/4}$. $\omega_{c}=\frac{4\pi}{15}$ and $t_{c}=t_{0}+D \cdot \exp(0.01t)$. Other parameters are consistent with those presented in Fig.~\ref{momentum oscillation}.\label{Kinetic energy oscillation}}
\end{center}
\end{figure}

%%%%
Fig.~\ref{Kinetic energy oscillation} show the dynamical behaviors of kinetic energy in the system. In the Hermitian system (i.e., $\lambda=0$) [see Fig.~\ref{Kinetic energy oscillation}(a)], the kinetic energy diffuses in a way of double-exponential function and finally reaches saturation, i.e., $\langle E_{k}\rangle=E_{ks}-A_{1}\exp(-\frac{t}{\mu_{1}})-A_{2}\exp(-\frac{t}{\mu_{2}})$. In the non-Hermitian system (i.e., $\lambda \neq 0$), the oscillation of kinetic energy occurs. Introducing a sufficiently strong non-Hermitian driving potential, e.g., $\lambda=0.5$ [see Fig.~\ref{Kinetic energy oscillation}(c)], the kinetic energy in the system oscillates as a damped cosine function,
\begin{equation}\label{kinetic}
  \langle E_{k} \rangle=E_{ka}(t)-E_{km}(t) \cdot \text{cos}\left[\omega_{c}(t-t_{c})\right]\;,
\end{equation}
where the asymptotic curve is $E_{ka}(t)=E_{ks}+B_{1}\cdot\exp(-\frac{t}{\mu})$, and the damped amplitude is $E_{km}(t) \propto \exp(-\frac{t}{\tau})$. Moreover, the oscillation frequency $\omega_{c}$ and the phase shift $t_{c}$ are identical to those of the corresponding momentum $\langle p\rangle$. With the increase of non-Hermitian driving strength, the decaying function of $E_{km}(t)$ also changes correspondingly, e.g., $E_{km}(t)\propto \exp\left(-\frac{t}{\tau}\right)(\frac{2}{\pi t})^{1/4}$ for $\lambda=1$ [see Fig.~\ref{Kinetic energy oscillation}(d)]. The research shows that the non-Hermitian driving also results in the damped oscillation of kinetic energy and can control its amplitude, phase shift, and asymptotic curve under resonant conditions.

%%%%
\begin{figure}[htbp]
\begin{center}
\includegraphics[width=7.8cm]{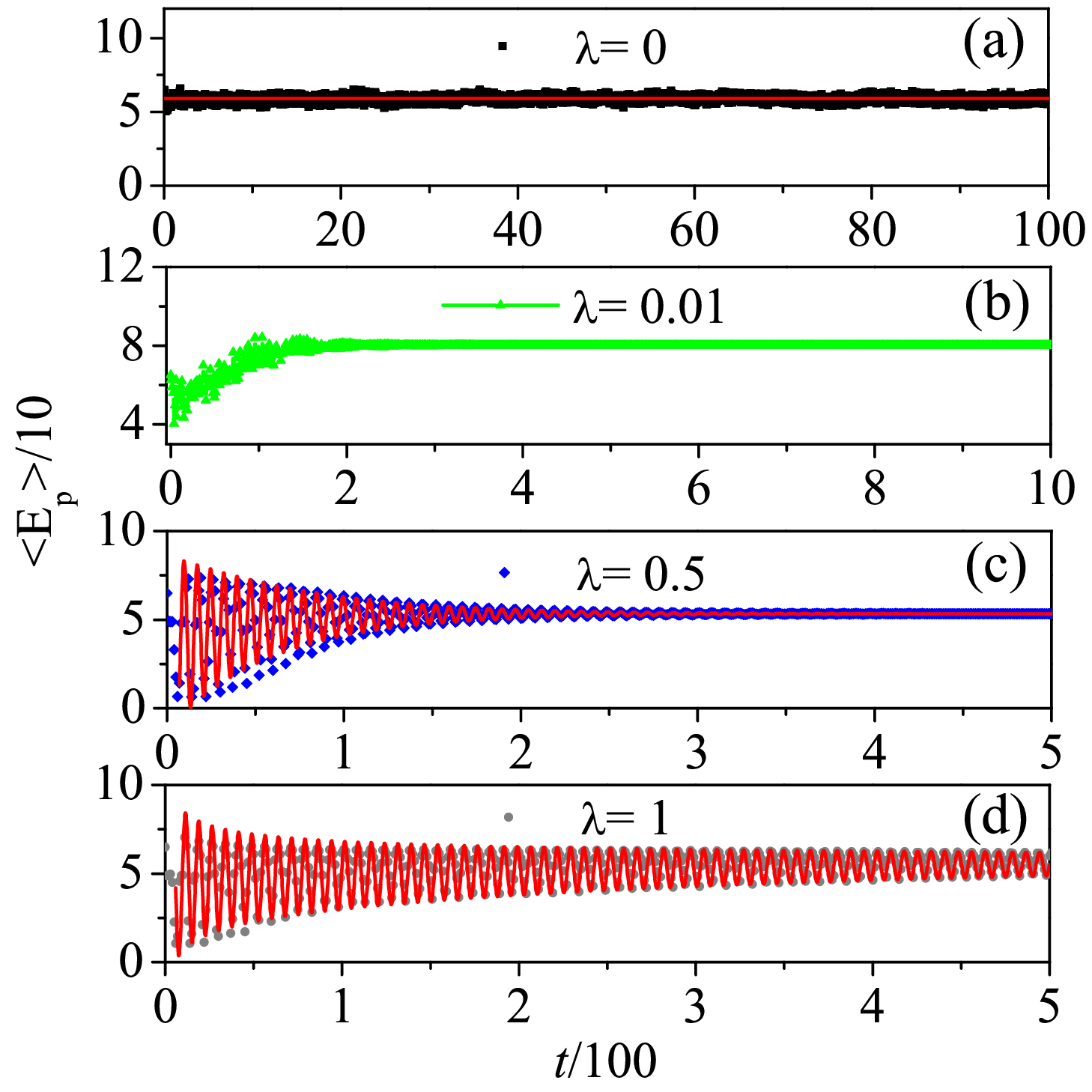}
\caption{Time evolution of $\langle E_{p} \rangle$ with $\lambda=0$ (squares) (a), 0.01 (triangles) (b), 0.5 (diamonds)(c), and 1 (circles) (d). Red line in (a) indicates $\langle E_{p}\rangle=C$; in (c) $\langle E_{p} \rangle=E_{pa}(t)+E_{pm}(t) \cdot \text{cos}\left[\omega_{c}(t-t_{c})\right]$ in Eq.~\eqref{potential} with $E_{pa}(t)=E_{ps}-17\cdot\exp(-\frac{t}{40})$ and $E_{pm}(t)=50\cdot \exp(-\frac{t}{66})$;
in (d) $\langle E_{p} \rangle=E_{pa}(t)+E_{pm}(t) \cdot \text{cos}\left[\omega_{c}(t-t_{c})\right]$ in Eq.~\eqref{potential} with $E_{pa}(t)=E_{ps}-12\cdot\exp(-\frac{t}{300})$ and $E_{pm}(t)=80\cdot \exp\left(-\frac{t}{600}\right)(\frac{2}{\pi t})^{1/4}$. $\omega_{c}=\frac{4\pi}{15}$ and $t_{c}=t_{0}+D \cdot \exp(0.01t)$. Other parameters are consistent with those presented in Fig.~\ref{momentum oscillation}.\label{Potential energy oscillation}}
\end{center}
\end{figure}

The dynamical behaviors of potential energy in the system are shown in Fig.~\ref{Potential energy oscillation}. In the Hermitian phase (i.e., $\lambda=0$) [see Fig.~\ref{Potential energy oscillation}(a)], the potential energy remains invariant throughout the time evolution process, i.e., $\langle E_{p}\rangle = C$. However, the potential energy oscillates with time in the non-Hermitian phase (i.e., $\lambda \neq 0$). With the sufficiently strong non-Hermitian driving, e.g., $\lambda=0.5$ [see Fig.~\ref{Potential energy oscillation}(c)], the potential energy oscillates as a damped cosine function,
\begin{equation}\label{potential}
  \langle E_{p} \rangle=E_{pa}(t)+E_{pm}(t)\cdot\text{cos}\left[\omega_{c}(t-t_{c})\right]\;.
\end{equation}
Here the asymptotic curve is $E_{pa}(t)=E_{ps}-B_{2}\cdot\exp(-\frac{t}{\mu})$ with the saturation $E_{ps}$, and the damped amplitude is $E_{pm}(t) \propto \exp(-\frac{t}{\tau})$. Increasing the driving strength $\lambda$, e.g., $\lambda=1$ [see Fig.~\ref{Potential energy oscillation}(d)], the amplitude decays as another function, i.e., $E_{pm}(t)\propto \exp\left(-\frac{t}{\tau}\right)(\frac{2}{\pi t})^{1/4}$. The oscillation frequency ($\omega_{c}$) of $\langle E_{p} \rangle$ is identical to that of the corresponding $\langle E_{k} \rangle$ but its phase leads $\pi$. This indicates that the kinetic energy $\langle E_{k} \rangle$ and the potential energy $\langle E_{p} \rangle$ mutually convert into one another during the time evolution, which is the characteristic feature of oscillator~\cite{JJahanpanah24}. Obviously, the non-Hermitian driving induces the damped oscillation in the potential energy as well and can adjust its amplitude, phase shift, and asymptotic curve under resonant conditions.

%%%%
%%%%%
Based on the discussion above, it is easy to calculate the total energy in the system, i.e., $\langle E \rangle=\langle E_{k} \rangle+\langle E_{p} \rangle$. In the Hermitian system (i.e., $\lambda= 0$),
the total energy diffuses as a double-exponential function, i.e., $\langle E\rangle=E_{s}-A_{1}\exp(-\frac{t}{\mu_{1}})-A_{2}\exp(-\frac{t}{\mu_{2}})$. In the non-Hermitian system (e.g., $\lambda =0.5$ and $\lambda =1$), however, the total energy oscillates as a damped cosine function,
\begin{equation}\label{energy1}
  \langle E \rangle=E_{a}(t)-E_{m}(t)\cdot\text{cos}\left[\omega_{c}(t-t_{c})\right]\;.
\end{equation}
Here the oscillation frequency $\omega_{c}$ and the phase shift $t_{c}$ are identical to those of the $\langle E_{k} \rangle$ and $\langle p \rangle$. The asymptotic curve is $E_{a}(t)=E_{s}+B_{3}\cdot\exp(-\frac{t}{\mu})$ with the saturation $E_{s}=E_{ks}+E_{ps}$ and the coefficient $B_{3}=B_{1}-B_{2}$, and the damped amplitude is $E_{m}(t)=E_{km}(t)-E_{pm}(t)$. The results show that there is a net increase in the total energy during the time evolution when subjected to non-Hermitian driving. In a word, the oscillations of momentum and energy indicate the oscillation of probability density of the system~\cite{GJi24}. It is evident that the PTQKHO system can be recognized as a damped harmonic oscillator with the fixed oscillation frequency $\omega_{c}$ under the resonant conditions. It is a oscillation mode reshaped by the resonant coupling between the non-Hermitian driving with the driving frequency $\Omega$ and the harmonic oscillator with the oscillation frequency $\omega$. It is reasonable to believe that the oscillation frequency $\omega_{c}$ is closely related to the ratio ($\omega/\Omega$) between the oscillation frequency $\omega$ and the driving frequency $\Omega$.

\section{Conclusion}\label{SEC-SUM}
In this work, we investigate the dynamical behaviors of momentum and energy in the non-Hermitian PTQKHO system. The results indicate that there are significant differences in dynamics between the non-resonant states and the resonant states. The nearest-neighbor hopping between momentum eigenstates with the non-Hermitian driving results in the existence of directed momentum current and ballistic energy diffusion under the non-resonant conditions. The PTQKHO system in this context can be characterized as a quasi-classical particle subjected to a constant non-Hermitian driving force. Under the resonant conditions, both the momentum and energy oscillate damply with the identical frequencies. The PTQKHO system in this context can be regarded as a damped harmonic oscillator that arises from the resonant coupling between the non-Hermitian driving and the harmonic oscillator. Our investigations reveal that the non-Hermiticity and the frequency characteristic of this system give rise to these distinctive dynamical behaviors together, which are useful in quantum control of ultra-cold atoms and electrons in solid state.

\textcolor{blue}{\textit{Acknowledgements}---}
This work is supported by the National Natural Science Foundation of China (Grant No.12364024), and the Dr. Start-up Fund of Jiangxi University of Science and Technology (No. 205200100067).

\end{document}